\documentclass[aps,twocolumn,groupedaddress,amsmath]{revtex4}
\usepackage{graphicx}
\begin{document}


\title{Precision of Quantization of the Hall Conductivity\\
in a Sample of Finite Size: Power Law}


\author{A.A.~Greshnov}
\email[]{a_greshnov@hotmail.com}
\author{G.G.~Zegrya}
\author{E.N.~Kolesnikova}
\affiliation{Ioffe Institute RAS, Polytekhnicheskaya st. 26, 194021, St.
Petersburg, Russia}


\date{\today}

\begin{abstract}
A microscopic calculation of the conductivity in the integer quantum Hall
effect (IQHE) regime is carried out. The problem of precision of quantization
is analyzed for samples of finite size. It is demonstrated that the precision
of quantization shows a power-law dependence on the sample size. A new scaling
parameter describing a dependence of this kind is introduced. It is also
demonstrated that the precision of quantization linearly depends on the ratio
between the amplitude of the chaotic potential and the cyclotron energy. The
results obtained are compared with the magnetotransport measurements in
mesoscopic samples.
\end{abstract}


\maketitle


 \subsection*{Introduction}
Despite the considerable progress in the understanding of the
quantum Hall effect (QHE), no consistent microscopic theory of
this phenomenon has been developed so far. It will be recalled
that the Hall resistance $R_H=h/\nu e^2$ is quantized in a strong
magnetic field directed perpendicularly to the plane of a
two-dimensional (2D) semiconductor sample~\cite{Kl}. Here $\nu$ is
an integer, and the precision of quantization at sufficiently low
temperatures is limited only by the measurement error, being as
good as a millionth of a percent~\cite{PrGir}. It is important
that the quantization occurs in a certain range of magnetic field
strengths or concentrations (on the Hall plateau). Such a behavior
of the conductivity of a 2D electron gas in a strong magnetic
field contradicts the results of the classical kinetic theory and
those obtained by the diagram technique of disorder-averaging
(e.g. in self-consistent Born approximation).

To describe IQHE, it is necessary to take into account the strong
localization of electrons in the chaotic impurity potential in a
magnetic field. The allowance for the chaotic potential leads to a
power-law dependence of the localization length $\xi$ on the
energy of the electronic state
$\xi\propto(E-E_n)^{-\nu},\,\nu\sim2.3$~\cite{bib:Huck}. The
existence of the scaling law for the energy dependence of the
localization length indicates that the chaotic potential should be
taken into account microscopically in order to calculate the
conductivity in the IQHE regime, which can be done using numerical
methods.

Previously, numerical calculations have been successfully applied
to study the dependence of the localization length on the energy
of electronic states. Several efficient methods are available,
which can be used to calculate the localization length for rather
large samples. Calculations of this kind have been employed to
confirm the theory of finite-size scaling and obtain values of
scaling indices that are in a good agreement with the
experiments~\cite{bib:Huck},\,\cite{bib:Koch}. Unfortunately,
these methods are inapplicable to calculations of the Hall
conductivity, because a complete knowledge of the carrier spectrum
and wave functions is necessary in this case.

In present paper, we calculate ab initio the Hall conductivity of
a 2D electron gas in a strong magnetic field. The results obtained
suggest that the precision of quantization of the Hall
conductivity on the plateau shows a power-law dependence on the
sample size and is directly proportional to the ratio between the
amplitude of the chaotic potential and the cyclotron energy.

Below we describe our model of the chaotic potential and the method used to
calculate the Hall conductivity. Then we present the results of our numerical
calculations and their analysis. Finally, we compare the theoretical results
with data obtained in magnetotransport measurements in mesoscopic samples.

 \subsection*{Model}
We consider a 2D electron gas at zero temperature $T=0$ and take
into account the elastic scattering of carriers only. Such an
approximation is justified under the following conditions. First,
the sample should not be too ''pure'', so that Coulomb effects
could not lead to transition to the regime of fractional QHE.
Second, a finite temperature results in the effective limitation
of the sample size by the phase-breaking length~$L_\phi\propto
T^{-p/2}$~\cite{bib:Thou}. Here, the scaling parameter $p$ is
determined by the prevalent mechanism of inelastic scattering and,
according to experimental data~\cite{bib:Koch}, is not a universal
parameter. Finally, the broadening of the Fermi step in the
distribution function can be disregarded because the typical
temperatures at which QHE is observed are low, $T<1K\simeq
0.1meV\ll\hbar\omega_c$.

The Hamiltonian of noninteracting carriers in the external
magnetic field $\bf B$ and the chaotic impurity potential $U({\bf
r})$ has the form:
\begin{eqnarray}
 \hat{H}_M=\frac{(\hat{\bf p}-e/c{\bf A})^2}{2m^*}+U(\bf r),\quad
 \mbox{rot }{\bf A}={\bf H}={\bf B}.
\end{eqnarray}

Let the magnetic field perpendicular to the plane of a 2D sample be directed
along the $z$-axis and let the chaotic potential $U({\bf r})$ to be independent
of $z$. We choose the vector potential of the uniform magnetic field in the
form ${\bf A}=(-By,0,0)$ (Landau gauge). When a single level in a quantum well
is considered, the problem is described by the following Hamiltonian:
\begin{eqnarray}
 \hat{H}=\frac{({\hat p}_x-eBy/c)^2+{\hat p}_y^2}{2m^*}+U({\bf r}),
 \quad {\bf r}=(x,y).
 \label{h0}
\end{eqnarray}

The model impurity potential used in this study has the following form:
\begin{eqnarray}
 U({\bf r})=\sum_{n=1}^{N}U_n
 \exp\left\{-\frac{({\bf r}-{\bf r}_n)^2}{R^2}\right\},
\end{eqnarray}
where the quantities $U_n$ and ${\bf r}_n$ are uniformly
distributed in the interval $[U_<,U_>]$ and over the entire plane
($x$,$y$), respectively. It makes possible to vary the amplitude
and the correlation properties of the potential with the use of
the parameters $N$, $U_<$, $U_>$ and $R$. At the same time, a
potential of this kind allows for analytical calculation of the
matrix elements in the basis of wave functions containing plane
waves~\cite{Landau}:
\begin{eqnarray}
\Psi_{n k}=\frac{\exp(ikx)}{\sqrt{2^n n!\sqrt{\pi}a_H L_x}}
\exp\left(-\frac{\widetilde{y}^2}{2a_H^2}\right)H_n
\left(\frac{\widetilde{y}}{a_H}\right),\label{eq2}\\
\widetilde{y}=y-ka_H^2,\quad a_H^2=\frac{\hbar c}{|e|B},\nonumber
\end{eqnarray}
where $n\geq0$ is a Landau level number, $H_n$ are the Hermite polynomials,
$a_H$ is a magnetic length. For a sample of finite dimensions $L_x \times L_y$,
the set of basis wave functions was determined by the conditions of periodicity
over a length $L_x$ and the point $y_0=ka_H^2$ falls within an interval of
length $L_y$~\cite{Landau}. The eigenenergies and wave functions needed to
calculate the conductivity can be found by numerical diagonalization of the
Hamiltonian $\hat H$ in the basis of the wave functions (\ref{eq2}).

An expression for the Hall conductivity (Kubo formula) can be derived at T = 0
using the first-order in electric field perturbation theory:
\begin{eqnarray}
\sigma_{xy}=\frac{e}{S}\sum_{\substack
{i<\mu\\f>\mu}}\frac{y_{if}(J_x)_{fi}+c.c.}
{\mathcal{E}_i-\mathcal{E}_f}, \label{Kubo}
\end{eqnarray}
where $\mathcal{E}_{i,f}$ are eigenenergies of $\hat H$, $y_{if}$ and
$(J_x)_{fi}$ are matrix elements of the coordinate and net current in the basis
of eigenfunctions of the Hamiltonian $\hat H$, $\mu$ is a chemical potential
and $S$ is a sample area.

 \subsection*{Results}
The above-described method for calculation of the Hall
conductivity of a 2D electron gas makes it possible to analyze the
influence of the finite sample size on the precision of
quantization. First of all, we note the importance of a consistent
calculation of both the density of states and the wave functions
of electronic states. Figure~1 shows the calculated density of
states $D({\mathcal E})$ and the typical electron density
distribution in localized and extended states. It can be seen that
the wave functions corresponding to the minimum density of states
are localized on a microscopic scale of the order of the magnetic
length, and those corresponding to the maximum density of states
are extended.

\begin{figure}[]
 \includegraphics[width=82mm]{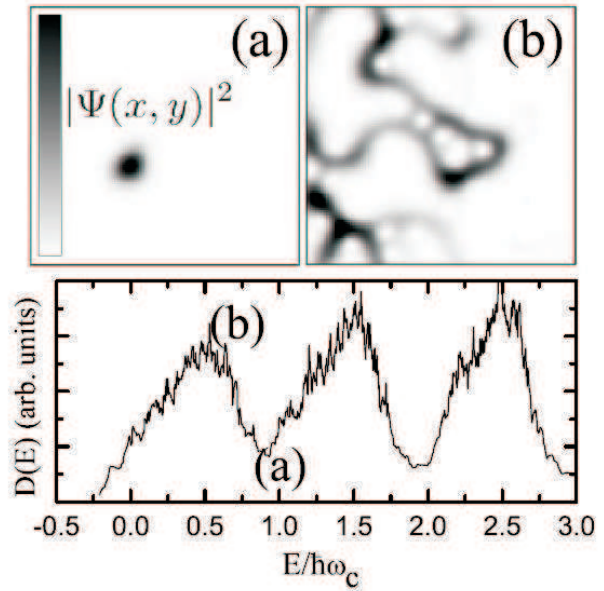}
\caption{Calculated density of states in a magnetic field (three lower Landau
levels are shown); distribution of electron density in (a) localized and (b)
extended states.}
\end{figure}

Figure~2 presents the results of Hall conductivity calculation for a sample
with such a dimensions that one Landau level contains 200 electron states. For
a magnetic field $B=10T$ this corresponds to a $0.3\times0.3\,\mu m^2$ sample.
The occurrence of strong conductivity fluctuations between QHE plateaus is a
mesoscopic effect intrinsic to the small samples ($\lesssim 1 \mu m$) at low
temperatures ($\lesssim 0.1~K$). For comparison with the results of our
calculations, Fig.~3 shows experimental data obtained in Ref.~\cite{bib:Cobd},
where magnetotransport measurements were carried out on a silicon MOSFET of
dimensions $0.6\times 0.6\, \mu m^2$ at a temperature of 100 mK. As can be seen
from Fig.~3, noticeable Hall conductivity fluctuations are observed on the
first and second QHE plateaus, in addition to those in the transition regions
between the plateaus.

\begin{figure}[]
 \includegraphics[width=82mm]{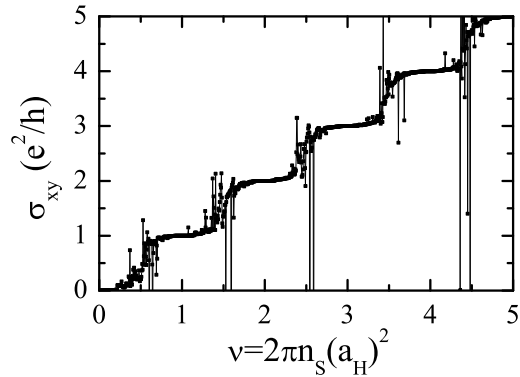}
\caption{Results of calculation of the Hall conductivity as a function of the
filling factor $\nu$.}
\end{figure}

\begin{figure}[]
 \includegraphics[width=82mm]{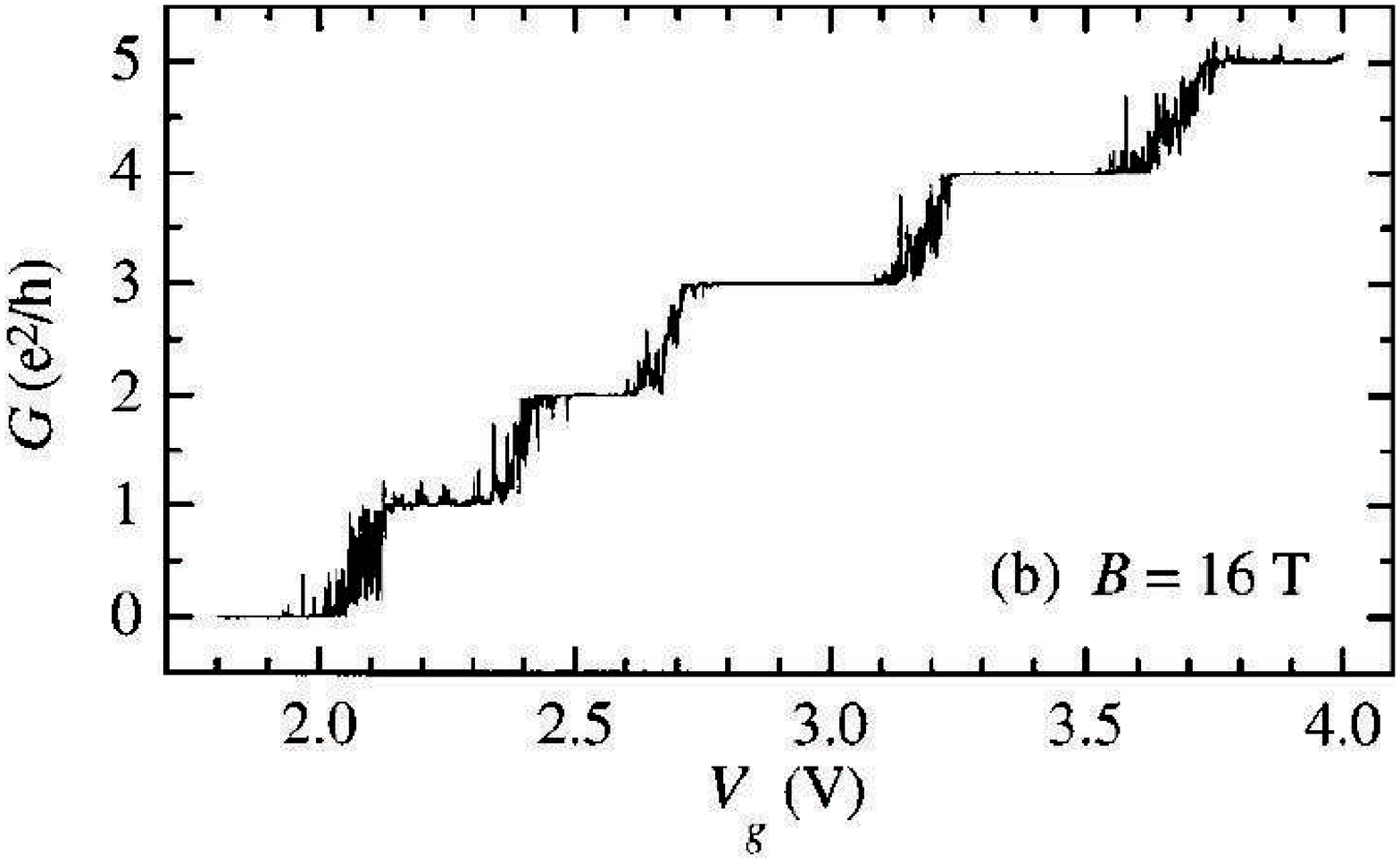}
\caption{Hall conductivity of a $0.6\times0.6\,\mu m^2$ silicon sample vs. the
gate voltage (reproduced from~\cite{bib:Cobd}).}
\end{figure}

To analyze the dependence of the precision of quantization on
sample size, we averaged deviations of the Hall conductivity of a
one filled Landau level from the quantized value $e^2/h$ over the
realizations of a random potential. The standard deviation of the
Hall conductivity at the center of the first plateau is plotted
against the sample size in Fig.~4. The presence in Fig.~4 of
equidistant straight lines in the double logarithmic scale means
that $\delta\sigma_{xy}$ is a power function of the ratio between
the sample size and the magnetic length and is proportional to the
amplitude of the chaotic potential $U_*$:
\begin{eqnarray}
 \frac{\delta\sigma_{xy}}{\sigma_{xy}}\propto
 \frac{U_*}{\hbar\omega_c}\left(\frac{a_H^2}{S}\right)^b.
 \label{eq8}
\end{eqnarray}
Here we introduce a new scaling parameter $b$, which describes
sample size dependence of the quantization precision. In all our
calculations, the parameter $b$ remains universal $b=0.7\pm5\%$.
The proportionality of $\delta\sigma_{xy}$ to the amplitude of the
chaotic potential is, in the limit $U_*\ll\hbar\omega_c$, an exact
analytical result, which is derived in the Appendix. As the
estimate based on Eq.~(\ref{eq8}) shows, fluctuations of the Hall
conductivity on the plateau may be as large as $10^{-2}$ for
samples of submicrometer dimensions. This estimation is in a
qualitative agreement with the experimental data
of~\cite{bib:Cobd}.

\begin{figure}[]
 \includegraphics[width=82mm]{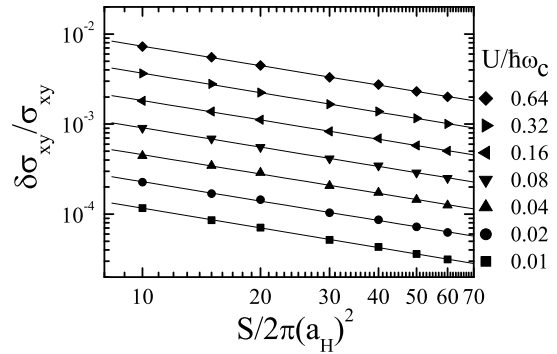}
\caption{Standard deviation of the Hall conductivity at the center of the first
plateau from $e^2/h$ as a function of the ratio between the sample area $S$ and
squared magnetic length $a_H^2$. The straight lines equidistant in the log--log
scale correspond to doubling of the amplitude of the chaotic potential (from
bottom to top).}
\end{figure}

It should be noted that the power-law dependence of the precision
of quantization on the sample size, obtained here, is
qualitatively confirmed by measurements performed on macroscopic
samples. In particular, Hall conductivity fluctuations have been
observed in the plateau at a level of $10^{-7}$ (at a measurement
error less than $10^{-8}$) for a silicon transistor of dimensions
$2\times 2\, mm^2$~\cite{Cage}. At the same time, most of the
theoretical studies known to us state that the correction to the
Hall conductivity, associated with the finiteness of the sample
size, is exponentially small~\cite{bib:Niu},~\cite{bib:Khm}.
Conclusions of this kind are based on the exponential behavior of
the wave functions of localized states. This, indeed, leads to an
exponentially small slope of the IQHE plateau, but does not
determine the precision with which the Hall conductivity in the
plateau takes a quantized value $\nu e^2/h$. The $\sigma_{xy}$
value on the plateau is determined by all electronic states lying
below the Fermi level, both localized and extended.

To conclude, we emphasize that the complete understanding of the
high precision quantization, observed in experiments on the
quantum Hall effect, requires further analysis of how the
precision of quantization depends on various factors. In the first
place, high-precision experiments where the dependence of the
precision of quantization on sample size, temperature, and carrier
mobility are to be performed. It is known from the temperature
dependence of the Hall conductivity of small ($L\sim10 \mu m$)
samples that, at temperatures $T\lesssim30mK$, the inelastic
scattering length can exceed the sample length
$L$~\cite{bib:Koch}. This means that such experimental conditions
open opportunities for direct observation of the deviation of the
Hall conductivity from the quantized value as a function of sample
size. It can be expected that measurements of this kind should
reveal a power-law, rather than exponential dependence of
$\delta\sigma_{xy}/\sigma_{xy}$ on $L$.

\subsection*{Acknowledgements}
The study was supported by the Russian Foundation for Basic Research
(04-07-90148, 04-02-16786, 05-02-16679) and by the Program for Support of
Leading Scientific Schools. One of the authors (A.A.G.) is grateful to the
Dynasty Foundation and the International Center for Fundamental Physics in
Moscow.

\subsection*{Appendix}
We now demonstrate that the Hall conductivity takes a quantized
value $\nu e^2/h$ at the center of each plateau in the limit
$U_*/\hbar\omega_c\rightarrow0$. Let the amplitude of the chaotic
potential $U_*$ be so small as compared to the cyclotron energy
$\hbar\omega_c$, that the density of states, $D({\mathcal E})$,
has an energy gap between Landau levels $N-1$ and $N$, $N\geq1$.
Let us calculate the Hall conductivity for the case when the
chemical potential lies within this gap, i.e., in the case of $N$
completely filled Landau levels.

If the condition $U_*\ll\hbar\omega_c$ is satisfied, the eigenenergies and
eigenfunctions of the Hamiltonian (\ref{h0}) can be found in terms of the
perturbation theory for degenerate states~\cite{Landau}. In this case, the
exact eigenenergies and wave functions can be represented as:
\begin{eqnarray}
{\mathcal E}_{n\alpha^{(n)}}=\hbar\omega_c(n+1/2+O(t)),\\
\Psi_{n\alpha^{(n)}}=\sum_k{}A_{\alpha^{(n)}k}\Psi_{nk}+
t\sum_{m\neq n, k}{}B_{\alpha^{(n)}mk}\Psi_{mk},\label{eq62}\\
\sum_k{A^*_{\alpha^{(n)}k}A_{\beta^{(n)}k}}=\delta_{\alpha^{(n)}\beta^{(n)}}+O(t).
\label{eq61}
\end{eqnarray}
Hereinafter we use Greek letters with superscripts $n(m)$ to denote energy
sublevels of the Landau level with a number $n(m)$: $\alpha^{(n)},\beta^{(m)}$;
$t$ is a small parameter $t=U_*/\hbar\omega_c$. The explicit form of the basis
functions $\Psi_{nk}$ is given by Eq.~(\ref{eq2}). The set of coefficients $A$
and $B$ depends on $t$, having a finite limit at $t\rightarrow0$; the
Eq.~(\ref{eq61}) is a direct consequence of the orthonormality of the set of
eigenfunctions $\Psi_{n\alpha^{(n)}}$. It is convenient to introduce auxiliary
wave functions
\begin{eqnarray}
\widetilde\Psi_{n\alpha^{(n)}}=\lim_{t\rightarrow0}\Psi_{n\alpha^{(n)}}=
\sum_k{}C_{\alpha^{(n)}k}\Psi_{nk} \label{eq63}
\end{eqnarray}
in such a way that they are exactly orthonormal and coincide with
the eigenfunctions $\Psi_{n\alpha^{(n)}}$ to within $O(t)$. A
consequence of the orthonormality of the wave functions
$\widetilde\Psi_{n\alpha^{(n)}}$ is the following identity derived
from Eq~(\ref{eq61}):
\begin{eqnarray}
\sum_k{C^*_{\alpha^{(n)}k}C_{\beta^{(n)}k}}=\delta_{\alpha^{(n)}\beta^{(n)}}.
\label{eq64}
\end{eqnarray}
Let us generalize the functions (\ref{eq63}) in the following way:
\begin{eqnarray}
\widetilde\Psi_{n\alpha^{(m)}}=\sum_k{}C_{\alpha^{(m)}k}\Psi_{nk}. \label{eq65}
\end{eqnarray}

Despite that at $m\neq n$ the wave functions $\widetilde\Psi_{n\alpha^{(m)}}$
are not eigenfunctions of the Hamiltonian $\hat H$ in the limit
$t\rightarrow0$, they are of use in further calculations. In particular,
$\widetilde\Psi_{n(\neq m)\alpha^{(m)}}$ can be expressed in terms of the wave
functions~(\ref{eq63}):
\begin{eqnarray}
\widetilde\Psi_{n\alpha^{(m)}}=\sum_{\beta^{(n)}}
D_{\alpha^{(m)}\beta^{(n)}}\widetilde\Psi_{n\beta^{(n)}}. \label{eq85}
\end{eqnarray}
Using Eqs.~(\ref{eq63})-(\ref{eq85}), we can readily obtain one more identity:
\begin{eqnarray}
\sum_{\gamma^{(n)}}D^*_{\alpha^{(m)}\gamma^{(n)}}D_{\beta^{(m)}\gamma^{(n)}}=
\delta_{\alpha^{(m)}\beta^{(m)}}.
\end{eqnarray}
To calculate the Hall conductivity by the Kubo formula~(\ref{Kubo}), it is
necessary, first of all, to derive expressions for the matrix elements of the
operators of the coordinate, $y$, and net current, $J_x$. In the basis of the
wave functions~(\ref{eq2}), we have:
\begin{eqnarray}
\langle\Psi_{nk}|\hat J_x|\Psi_{mq}\rangle=
\frac{e\hbar}{ma_H}\langle\Psi_{nk}|\hat\Lambda|\Psi_{mq}\rangle,\\
\langle\Psi_{nk}|y|\Psi_{mq}\rangle=a_H\langle\Psi_{nk}|\hat\Lambda|\Psi_{mq}\rangle+
ka_H^2\delta_{kq}\delta_{mn},\\
\langle\Psi_{nk}|\hat\Lambda|\Psi_{mq}\rangle=\sqrt{\frac{\max(m,n)}{2}}\delta_{kq}\delta_{|m-n|,1}.
\end{eqnarray}
Using the above identities, we obtain the following matrix elements between the
neighboring Landau levels:
\begin{eqnarray}
\langle\widetilde\Psi_{n-1,\alpha^{(n)}}|\hat\Lambda|\widetilde\Psi_{n\beta^{(n)}}\rangle=
\sqrt{{n}/{2}}\delta_{\alpha^{(n)}\beta^{(n)}}, \label{eq66}\\
\langle\widetilde\Psi_{n-1,\alpha^{(n-1)}}|\hat\Lambda|\widetilde\Psi_{n\beta^{(n)}}\rangle=
\sqrt{{n}/{2}}D_{\alpha^{(n-1)}\beta^{(n)}} \label{eq67}.
\end{eqnarray}
In Eqs. (\ref{eq66}), (\ref{eq67}) and below the operator $\hat\Lambda$ is
given by any of two following expressions:
\begin{eqnarray}
\hat\Lambda\equiv\{y/a_H,\,ma_H{\hat J_x}/(e\hbar)\}.
\end{eqnarray}
Thus, if the chemical potential level $\mu$ lies within the energy gap between
two Landau levels, calculation of the Hall conductivity by the Kubo formula
gives the following result:
\begin{widetext}
\begin{multline}
 \sigma^N_{xy}=\sum_{\substack{m\leq N-1,\,\alpha^{(m)} \\ n\geq N,\,\beta^{(n)}}}
 \frac{2e}{S(\mathcal{E}_{m\alpha^{(m)}}-\mathcal{E}_{n\beta^{(n)}})}
 \Re\langle\Psi_{m\alpha^{(m)}}|y|\Psi_{n\beta^{(n)}}\rangle
 \langle\Psi_{n\beta^{(n)}}|{\hat J_x}|\Psi_{m\alpha^{(m)}}\rangle=\\
 =\frac{2e^2}{\hbar}\frac{a_H^2}{S} \sum_{\substack{\alpha^{(N-1)} \\ \beta^{(N)}}}
 |\langle\widetilde\Psi_{N-1,\alpha^{(N-1)}}|\hat\Lambda|\widetilde\Psi_{N\beta^{(N)}}\rangle|^2+O(t)=
 \frac{Ne^2}{\hbar}\frac{a_H^2}{S}\sum_{\substack{\alpha^{(N-1)} \\ \beta^{(N)}}}
 D^*_{\beta^{(N)}\alpha^{(N-1)}} D_{\beta^{(N)}\alpha^{(N-1)}}+O(t)=\\
 =\frac{Ne^2}{\hbar}\frac{a_H^2}{S}\sum_{\beta^{(N)}}1+O(t)=\frac{Ne^2}{h}+O(t)
\end{multline}
\end{widetext}
Here the summation over $\beta^{(N)}$ reduces to calculation of the number of
states per Landau level in a sample of area S, and, therefore,
$\sum_{\beta^{(N)}}1=S/(2\pi a_H^2)$~\cite{Landau}.

Thus, we proved that the Hall conductivity at the center of a plateau
$\sigma_{xy}=\nu e^2/h+O(U_*/\hbar\omega_c)$.

\end{document}